\documentclass[aps,twocolumn,showpacs]{revtex4}
\usepackage{amsmath}
\usepackage{amssymb}
\usepackage{bm}
\usepackage{epsfig}
\usepackage{graphicx}
\usepackage{natbib}
\usepackage{color}
\usepackage[T2A]{fontenc}

\newcount\timehh  \newcount\timemm
\timehh=\time \divide\timehh by 60
\timemm=\time
\usepackage[cp1251]{inputenc}

\begin{document}

\title{Spin injection via (110)-grown semiconductor barriers} 
\author{P.\,S.\,Alekseev, M.\,M.\,Glazov, and S.\,A.\,Tarasenko}

\affiliation{A.\,F.\,Ioffe Physical-Technical Institute, 194021 St.-Petersburg, Russia}

\begin{abstract}
We study the tunneling of conduction electrons through a (110)-oriented single-barrier heterostructure
grown from III-V semiconductor compounds. It is shown that, due to low spatial symmetry of such a barrier, the tunneling current through the barrier leads to an electron spin polarization. The inverse effect, generation of a direct tunneling current by spin polarized electrons, is also predicted. We develop the microscopic theory of the effects and show that the spin polarization emerges due to the combined action of the Dresselhaus spin-orbit coupling within the barrier and the Rashba spin-orbit coupling at the barrier interfaces.
\end{abstract}


\maketitle

\section{Introduction}

The tunnel barriers play an important role in solid state physics. They determine the energy spectrum
of coupled quantum wells and quantum dots, stability of charged impurities and excitons in electric field,
miniband transport and Bloch oscillations in superlattices, etc., and underlie 
a number of electronic devices such as tunnel and resonant tunnel diodes~\cite{IvchenkoPikus_book,Davies_book}. The tunnel barrier is an essential ingredient of a ferromagnet-semiconductor
spin-injecting heterostructure which enables one to overcome the problem of the conductivity mismatch between the
ferromagnet and semiconductor and achieve a high efficiency of spin injection~\cite{Schmidt00,Rashba00,Zhu2001,Motsnyi03,Moser2007,Erve07,Salis2011,Han2013,Wang2013,Nestoklon2013}.
About a decade ago, it was realized that the process of electron tunneling is spin dependent itself due 
spin-orbit interaction which couples spin states with the orbital motion. It was shown that the Rashba spin-orbit 
coupling at interfaces as well as the Dresselhaus coupling in the bulk of barrier make the barrier tunnel transmission
dependent of the spin orientation and wave vector of incident electrons~\cite{Voskoboynikov1999,Silva1999,Perel2003,Zakharova2005}. The effect of spin-dependent tunneling was proposed to be applied for the pure electric injection and detection of spin polarized carriers~\cite{Hall2003,Tarasenko04,Khodas2004,Glazov2005,Mishra2005,Alekseev2006,Rozhansky2006,Sandu2006,Nguyen2009,Tkach2009,Leyva2010,Litvinov2010,Sherman2013}. In the case of Rashba and Dresselhaus coupling in (001)-grown barriers in bulk semiconductors, the spin polarization of transmitted electrons linearly scales with the lateral component $\bm{k}_{\parallel}$ of the electron wave vector and is of opposite
sign for the wave vectors $\bm{k}_{\parallel}$ and $-\bm{k}_{\parallel}$~\cite{Voskoboynikov1999,Perel2003}. Therefore, the proposed methods of spin injection require the application of an additional, lateral, electric field which causes the electron drift in the interface plane and makes the electron distribution in the $\bm{k}_{\parallel}$ space anisotropic.

In this paper, we show that the spin injection via tunnel structures in bulk semiconductors can be achieved even for isotropic 
in the interface plane electron distribution. Such an effect occurs for barriers of sufficiently low spatial symmetry only being not allowed, e.g., in (001)-grown structures. Microscopically, it is caused by the combined action of the Dresselhaus spin-orbit coupling in the barrier and the Rashba spin-orbit coupling at the barrier interfaces. To be specific, we consider single-barrier tunnel structures with the (110) crystallographic orientation and present an analytical theory of the spin injection.

\section{Microscopic Model}

Consider a zinc-blende-type semiconductor heretostructure with the symmetric potential barrier grown along the $z \parallel [110]$ axis,
see Fig.~1. 
\begin{figure}[b]
\includegraphics[width=0.9\linewidth]{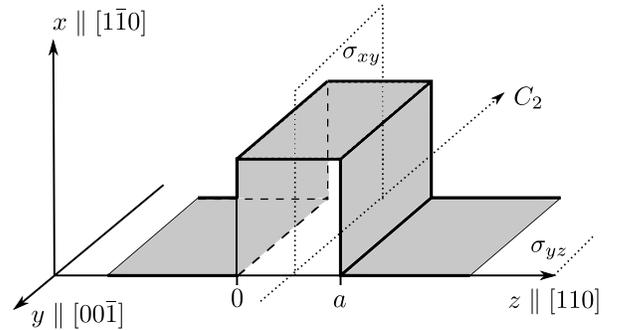}
\caption{\label{fig:scheme} The sketch of a symmetric tunnel barrier with the (110) crystallographic orientation. The point-group symmetry elements of the structure includes the two-fold rotation axis $C_2 \parallel y$ and the mirror planes $\sigma_{xy}$ and $\sigma_{yz}$.}
\end{figure}
Electrons with different incident wave vectors $\bm{k}=(k_x,k_y,k_z)$ tunnel through the barrier, where $k_x$ and $k_y$ are
projections of the wave vector onto the in-plane axes $x \parallel [1\bar{1}0]$ and $y \parallel [00\bar{1}]$. The effective Hamiltonian describing electron states in the conduction band has the form 
\begin{equation}\label{Htot}
\mathcal H = \mathcal H_0 + \mathcal H_{\rm R} + \mathcal H_{\rm D} \:,
\end{equation}
where $\mathcal H_0$ is the spin-independent contribution,
\begin{equation}\label{H0}
\mathcal H_0 = \frac{\hbar^2 (k_x^2+k_y^2)}{2m(z)} - \frac{\hbar^2}{2} \frac{\partial}{\partial z} \frac{1}{m(z)} \frac{\partial}{\partial z} +V(z) \:,
\end{equation}
$m(z)$ is the effective mass, which may be different inside and outside the barrier, $V(z)$ is the heteropotential, 
$V(z) = V_0>0$ if $0<z<a$ and $V(z)=0$ elsewhere, $a$ is the barrier thickness, $\mathcal H_{\rm R}$ is the Rashba spin-orbit coupling at interfaces
\begin{equation}\label{H_R}
 \mathcal{H}_{\mathrm{R}} = \alpha \left[ \delta(z-a) - \delta(z) \right] ( \sigma_x k_y - \sigma_y k_x ) \:,
\end{equation}
$\alpha$ is the parameter determined by the band offsets, $\sigma_j$ ($j=x,y,z$) are the Pauli matrices, and $\mathcal H_{\rm D}$ is the Dresselhaus spin-orbit coupling. In the chosen coordinate frame, $\mathcal H_{\rm D}$ can be presented as the sum of four terms
\begin{equation}\label{H_D1}
\mathcal H_{D1} = i \frac{\sigma_x}{2}  \left\{ \gamma(z),
\frac{\partial^3}{\partial z^3}\right\}_{\mathrm{sym}}\:,
\end{equation}
\begin{equation}\label{H_D2}
\mathcal H_{D2} = \frac{\sigma_z k_x }{2} \frac{\partial}{\partial z} \gamma(z) \frac{\partial}{\partial z} \:,
\end{equation}
\begin{equation}\label{H_D3}
\mathcal H_{D3} = i \left[ \sigma_x \left( \frac{k_x^2}{2} + k_y^2\right) -2 \sigma_y k_x k_y \right] \left\{ \gamma(z), \frac{\partial}{\partial z} \right\}_{\mathrm{sym}} \:,
\end{equation}
\begin{equation}\label{H_D4}
\mathcal H_{D4} = \sigma_z k_x \left( \frac{k_x^2}{2} -k_y^2 \right) \gamma(z)  \:,
\end{equation}
where $\gamma(z)$ is the bulk Dresselhaus parameter and the notation $\{ \ldots \}_{\rm sym}$ denotes the operator symmetrization.
The terms $\mathcal H_{D1}$, $\mathcal H_{D2}$, $\mathcal H_{D3}$, and $\mathcal H_{D4}$ contain the derivative of the third, second, first, and zero order, respectively. We assume the kinetic energy of electrons to be substantially smaller than the barrier height $V_0$ and, therefore, neglect the terms $\mathcal H_{D3}$ and $\mathcal H_{D4}$ in comparison with $\mathcal H_{D1}$ and $\mathcal H_{D2}$, respectively. 

We are interested in the total spin polarization proportional to the tunneling current though the barrier $j_z$. In (110)-grown structures, such an effect is allowed by symmetry and phenomenologically given by
\begin{equation}\label{phen}
s_x(z) \propto j_z \:,
\end{equation}
where $s_x(z)$ is the steady-state spin density.
Indeed, symmetric (110)-grown barriers are described by the $C_{2v}$ point group which consists of the two-fold rotation axis $C_2 \parallel y$, two mirror planes $\sigma_{xy} \parallel (110)$ and $\sigma_{yz} \parallel (1\bar{1}0)$, and the identity element~\cite{Nestoklon2012b}, as shown in Fig.~\ref{fig:scheme}. In this point group, the vector component $j_z$ and pseudovector component $s_x$ belong to the same irreducible representation, i.e., under all symmetry operations of the group they transform in a similar way~\cite{Olbrich2009}. Therefore, the symmetry allows the linear coupling of $s_x$ and $j_z$ and imposes the condition that the current-induced spin density $s_x(z)$ is an even function with respect to the barrier center. The latter suggests that the electrons transmitted through the barrier and those reflected from the barrier have the same spin polarization. Note that, in (001)-grown structures, the spin injection is absent for the tunneling current along the barrier normal and may occur only in the presence of the lateral component of the current. Equation~\eqref{phen} also shows the possibility of the inverse effect: A non-equilibrium spin polarization of carriers along the $x$ axis causes the direct current through the barrier.

We assume that the electrons incident upon the barrier are unpolarized and their distribution in the interface plane is isotropic.
Symmetry analysis of the spin-orbit terms shows that the spin injection given by Eq.~\eqref{phen} can be microscopically related to (i) joint action of the Rashba $\mathcal H_{R}$ and Dresselhaus $\mathcal H_{D2}$ terms or (ii) solely $\mathcal H_{D1}$ term. The calculation based on the perturbation theory demonstrates that the term $\mathcal H_{D1}$ does not lead to a spin injection, see Appendix and Ref.~\cite{Nguyen2009}. Therefore, we focus below on the mechanism caused by the joint action of the Rashba spin-orbit coupling at interfaces and Dresselhaus coupling in the barrier bulk. 

\begin{figure}[tb]
\includegraphics[width=0.7\linewidth]{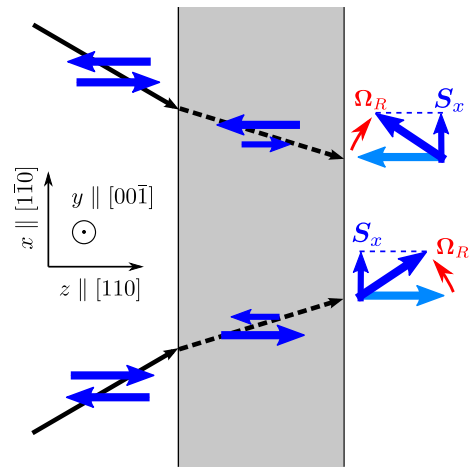}
\caption{\label{fig:micro} 
(Color online) The model of spin injection via (110)-grown barrier.
The spin component $S_x>0$ of electrons transmitted through the barrier with different in-plane wave vectors emerges due to (i) anisotropic spin filtering caused by the Dresselhaus spin-orbit coupling in the barrier interior followed by (ii) spin rotation in the interface-induced Rashba effective magnetic field $\bm \Omega_R$. }
\end{figure}

The mechanism of the spin injection can be viewed as a two-step process illustrated in Fig.~\ref{fig:micro}. At the first stage, unpolarized electrons with different
in-plane wave vectors $k_x$ are incident upon the barrier and tunnel through it being affected by the Dresselhaus spin-orbit coupling. The term $\mathcal H_{D2}$ can be considered as the spin-dependent correction to the effective mass along $z$ which is proportional to $k_x$, compare Eqs.~\eqref{H0} and~\eqref{H_D2}. Due to the fact that the barrier transmission probability depends on the effective mass, the electrons at the exit from the barrier gain the spin polarization 
$S_z \propto k_x$. Similar mechanism of anisotropic spin filtering caused by the Dresselhaus coupling for (001)-grown structures was considered in Refs.~\cite{Perel2003,Glazov2005}. Note that, for the equal population of the $k_x$ and $-k_x$ states, the net spin polarization of electrons is absent at this step.
The net spin polarization emerges at the second stage due to the action of the Rashba spin-orbit coupling at the interface, see Eq.~\eqref{H_R}. The Rashba coupling can be considered as an effective magnetic field $\bm \Omega_R$ lying in the interface plane which causes the spin rotation. The $y$-component of the effective field $\bm \Omega_R$ is proportional to $k_x$, therefore, the rotation axes are opposite for the electrons with positive and negative $k_x$. As a result, the rotation
leads to the spin component $S_x>0$ for all the electrons transmitted though the barrier. The value of the net spin polarization is determined by the
efficiency of anisotropic spin filtering in the barrier bulk and the spin rotation angle at the interface. We note that similar mechanism leads to the spin polarization of electrons reflected from the barrier.

\section{Theory}

The theory of spin injection can be conveniently developed by using the spin-dependent transfer matrix technique. The $4\times4$ transfer matrix 
$\mathcal{P}$ through the whole tunnel structure relates the spinor wave function $\Psi$ and its derivative $\Psi'=d \Psi /dz$ at $z=-0$ to those at $z=a+0$,
\begin{equation}\label{P}
\left(
\begin{array}{c}
\Psi(a+0) \\ \Psi'(a+0)
\end{array}
\right) = \mathcal{P} \left(
\begin{array}{c}
\Psi(-0) \\ \Psi'(-0)
\end{array}
\right) \:, \;\;  \mathcal{P} = \left(
\begin{array}{cc}
P_{\psi\psi} & P_{\psi\psi'} \\ P_{\psi'\psi} & P_{\psi'\psi'}
\end{array}
\right) \:,
\end{equation}
where $P_{\psi\psi}$, $P_{\psi\psi'}$, $P_{\psi'\psi}$, and $P_{\psi'\psi'}$ are $2\times2$ matrices. Higher-order derivatives of the wave function are expressed via 
$\Psi$ and $\Psi'$ because the Schr\"{o}dinger equation with the Hamiltonian $\mathcal H_0 + \mathcal H_{D2} + \mathcal H_{R}$ is a differential equation of the second order in $z$. By definition, the matrix $\mathcal{P}$ can be presented as the ordered product
of the transfer matrices through individual parts of the structure,
\begin{equation}\label{Ptot}
\mathcal P = \mathcal P^{(r)} \mathcal P^{(b)} \mathcal P^{(l)}, 
\end{equation}
where $\mathcal P^{(l/r)}$ is the transfer matrix through the left/right interface and $\mathcal P^{(b)}$ is the transfer matrix through the barrier interior. 

Solution of the Schr\"{o}dinger equation at the left interface with the standard boundary conditions, which follow from the Hamiltonian $\mathcal H_0 + \mathcal H_{D2} + \mathcal H_{R}$ and require the continuity of $\Psi$ and $(1/m - \gamma \sigma_z k_x /\hbar^2)\Psi'$, shows that the transfer matrix $\mathcal P^{(l)}$ has the components
\begin{subequations}
\label{transfer:int}
\begin{equation}\label{Ppp:int:110}
P_{\psi\psi}^{(l)} = I \:, \qquad
P_{\psi\psi'}^{(l)} = 0 \:,
\end{equation}
\begin{equation}\label{Pp1p:int:110}
P_{\psi'\psi}^{(l)} = -\frac{2m_2 {\alpha}}{\hbar^2} 
\left(I- \frac{m_2\gamma_2}{\hbar^2} \sigma_z  k_x
\right)^{-1} \left(\sigma_x k_y - \sigma_y k_x \right) ,
\end{equation}
\begin{equation}\label{Pp1p1:int:110}
P_{\psi'\psi'}^{(l)} =
\frac{m_2}{m_1} \left(I- \frac{m_2\gamma_2}{\hbar^2} \sigma_z k_x
\right)^{-1} \left(I - \frac{m_1\gamma_1}{\hbar^2} \sigma_z k_x 
\right) ,
\end{equation}
\end{subequations}
where $m_2$ and $\gamma_2$ are the effective mass and Dresselhaus constant inside the barrier, $m_1$ and $\gamma_1$ are those outside the barrier, and $I$ is the $2\times 2$ unit matrix. The transfer matrix through the right interface $\mathcal P^{(r)}$ is equal to the inverse transfer matrix through the left interface, $\mathcal P^{(l)-1}$, and readily obtained from Eqs.~\eqref{transfer:int} by the replacement $\alpha \to - \alpha$, $m_{1(2)} \to m_{2(1)}$, and $\gamma_{1(2)} \to \gamma_{2(1)}$. The transfer matrix through the barrier interior with the term $\mathcal H_{D2}$ included in the Hamiltonian is given by
\begin{subequations}
\begin{equation}\label{Ppp:110}
P_{\psi\psi}^{(b)} = P_{\psi'\psi'}^{(b)} =  \left(%
\begin{array}{cc}
  \cosh{q_+ a} & 0 \\
  0 & \cosh{q_- a} \\
\end{array}%
\right) ,
\end{equation}
\begin{equation}\label{Ppp1:110}
P_{\psi\psi'}^{(b)} = \left(%
\begin{array}{cc}
  q_+^{-1}\sinh{q_+ a} & 0 \\
  0 & q_-^{-1}\sinh{q_- a} \\
\end{array}%
\right) ,
\end{equation}
\begin{equation}\label{Pp1p:110}
P_{\psi'\psi}^{(b)} = \left(%
\begin{array}{cc}
  q_+\sinh{q_+ a} & 0 \\
  0 & q_-\sinh{q_- a} \\
\end{array}%
\right),
\end{equation}
\end{subequations}
where 
\begin{equation}\label{qpm}
q_{\pm} = q {\left({1\pm \frac{\gamma_2  m_2
k_x}{\hbar^2}}\right)^{-1/2}} 
\end{equation}
and
\begin{equation}\label{q0}
q = \sqrt{\frac{2m_2V_0}{\hbar^2} - k_z^2\frac{m_2}{m_1} -
k_\parallel^2\left(\frac{m_2}{m_1}-1\right)}
\end{equation}
with $k_\parallel = \sqrt{k_x^2+k_y^2}$ being the in-plane wave vector. The inverse matrix $\mathcal P^{(b)-1}$ is obtained from $\mathcal P^{(b)}$ by the replacement $a \rightarrow -a$.

The knowledge of the transfer matrix $\mathcal P$ allows one to calculate the spin-dependent coefficient of electron transmission and reflection from the tunnel structure. Assuming that the spin-orbit interaction is negligible outside the barrier and electrons are incident upon the structure from the left, we present the wave function describing the incident, reflected, and transmitted waves in the form 
\begin{equation}\label{psi}
\psi_{\bm{k}}(\bm{r}) = \left\{
\begin{array}{l}
\chi \, e^{i \bm k_{\parallel} \cdot \bm{\rho} \,+\, ik_z z} + \mathcal R_{\bm k} \chi \: e^{i \bm k_{\parallel} \cdot \bm{\rho} \,-\, ik_z z} \:,\;\; z<0 \:, \\
\mathcal T_{\bm k} \chi \, e^{i \bm k_{\parallel} \cdot \bm{\rho} \,+\, i k_z (z-a)} \:, \;\; z>a 
\end{array} \right. 
\end{equation}
where $\chi$ is the incident electron spinor, $\mathcal R_{\bm k}$  and $\mathcal T_{\bm k}$ and are $2 \times 2$ matrices of amplitude reflection and transmission coefficients. It follows from Eqs.~\eqref{P} and~\eqref{psi} that the matrices $\mathcal R_{\bm k}$  and $\mathcal T_{\bm k}$ satisfy the equation
\begin{equation}
\label{transfer:tr}
\begin{pmatrix}
\mathcal T_{\bm k} \\
i k_z \mathcal T_{\bm k}
\end{pmatrix}
=
\mathcal P 
\begin{pmatrix}
I + \mathcal R_{\bm k}\\
i k_z (I - \mathcal R_{\bm k})
\end{pmatrix} ,
\end{equation}
which yields
\begin{equation}
\label{refl}
\mathcal R_{\bm k} = - \left( \mathcal P_{\psi\psi} + \mathcal P_{\psi'\psi'} - i k_z  \mathcal P_{\psi \psi'} + i k_z^{-1}  \mathcal P_{\psi'\psi} \right)^{-1} 
\end{equation}
\[
\times \left( \mathcal P_{\psi\psi} -  \mathcal P_{\psi'\psi'} + i k_z \mathcal P_{\psi\psi'} + i k_z^{-1} \mathcal P_{\psi'\psi} \right) ,
\]
\begin{equation}
\label{transm}
\mathcal T_{\bm k} = 2 \left( \bar{\mathcal P}_{\psi\psi} + \bar{\mathcal P}_{\psi'\psi'} + i k_z \bar{\mathcal P}_{\psi \psi'} - ik_z^{-1} \bar{\mathcal P}_{\psi'\psi}  \right)^{-1} ,
\end{equation}
where $\bar{\mathcal P} = \mathcal P^{-1}$, i.e., $\bar{\mathcal P}_{\psi\psi}$, $\bar{\mathcal P}_{\psi\psi'}$, $\bar{\mathcal P}_{\psi'\psi}$, and $\bar{\mathcal P}_{\psi'\psi'}$ are the $2\times 2$ blocks of the inverse transfer matrix. Since $\mathcal P = \mathcal P^{(r)} \mathcal P^{(b)} \mathcal P^{(l)}$, $\mathcal P^{(l)-1} = \mathcal P^{(r)}$ , $\mathcal P^{(r)-1} = \mathcal P^{(l)}$, and $\mathcal P^{(b)-1}(a) = \mathcal P^{(b)}(-a)$, the inverse transfer matrix through the structures is given by $\mathcal P^{-1}(a) = \mathcal P(-a)$. It follows that the amplitude coefficients of transmission and reflection for the electrons incident upon the barrier from the right (with $k_z <0$) are obtained from $\mathcal T_{\bm k}$ and $\mathcal R_{\bm k}$, respectively, by the replacement $a \rightarrow -a$.

\subsection{Spin injection}

The spin of electrons with the wave vector $\bm k$ transmitted through the barrier is given by 
\begin{equation}
\label{Sk}
\bm S_{\bm k} = \frac{{\rm Tr} (\bm{\sigma} g_{\bm{k}})}{2 {\rm Tr} (g_{\bm{k}})} \:,
\end{equation}
where $g_{\bm{k}}$ is the $2\times2$ spin matrix describing the charge and spin fluxes of the transmitted particles. In the case when the electrons incident upon the barrier are unpolarized, the matrix $g_{\bm{k}}$ has the form~\cite{Tarasenko04}
\begin{equation}\label{g:mat}
g_{\bm k} = \mathcal T_{\bm k} \mathcal T_{\bm k}^{\dag} f_{\bm{k}} v_z \Theta(v_z) \:,
\end{equation}
where $f_{\bm{k}}$ is the distribution function of incident electrons, $v_z = \hbar k_z /m_1$ is the velocity component along the barrier normal, and $\Theta(v_z)$ is the Heaviside step function. Note that the matrix describing the flux of reflected electrons (with $k_z <0$) is obtained from Eq.~\eqref{g:mat} by the replacement $\mathcal T_{\bm k} \to \mathcal R_{\bm k}$ and $k_z \to -k_z$.

The equations presented above enable one to calculate the spin polarization of transmitted and reflected electrons for given parameters of the tunnel structure. To simplify the calculation, we assume that the effective masses inside and outside the barrier are the same ($m_1 = m_2 =m$) and neglect the spin-orbit coupling outside the barrier ($\gamma_2 = \gamma$, $\gamma_1=0$). In this case, to first order in $\alpha$, $\gamma$, and $\alpha \gamma$, the matrix $\mathcal T_{\bm k}$ has the form
\begin{equation}\label{Tk:res}
\mathcal T_{\bm k} = \left[I + \frac{\gamma m c_1 }{4 \hbar^2} \sigma_z k_x + \frac{\alpha\gamma m^2 c_2}{\hbar^4 q} (\sigma_x k_x + \sigma_y k_y)k_x \right] t_{\bm k} ,
\end{equation}
where $c_1$ and $c_2$ are the coefficients given by
\begin{subequations}
\begin{multline}
\label{C1}
c_1 = \left[ i q a \left(k_z/q - q/k_z \right) \cosh{q a} \right.\\
 - \left. \left(2q a + 3k_z/q + 3q/k_z \right) \sinh{q a} \right] t_{\bm k},
\end{multline}
\begin{equation}
\label{C2}
c_2 = [\left( 3i + q^2 a /k_z \right) \sinh{q a} - i q a \cosh{q a} ]\, t_{\bm k} \:,
\end{equation}
\end{subequations}
and $t_{\bm k}$ is the amplitude transmission coefficient in the absence of spin-orbit coupling,
\begin{equation}\label{tk:res}
t_{\bm k} = \left[ \cosh{q a} + \frac{i}{2}\left(\frac{q}{k_z}-
\frac{k_z}{q}\right) \sinh{q a} \right]^{-1} \:.
\end{equation}
Then, the symmetric part of the spin distribution of transmitted electrons assumes the form
\begin{equation}\label{S_x}
S_{\bm k,x}^{({\rm sym})} = 2 \frac{\alpha \gamma m^2 k_x^2 k_z}{\hbar^4} 
\end{equation}
\vspace{-0.5cm}
\[
\times \frac{3(q^2-k_z^2)\sinh^2 q a + q a (k_z^2+q_0^2) \cosh q a \sinh q a}{(k_z^2+q^2)^2\cosh^2 qa 
- (k_z^2 - q^2)^2} .
\]
In the particular case of high and thick enough barrier, $qa \gg 1$ and $q \gg k_z$, Eq.~\eqref{S_x} is simplified to
\begin{equation}\label{Sx:deep}
S_{\bm k,x}^{({\rm sym})} = 2 \frac{\alpha\gamma m^2 k_x^2 k_z a }{\hbar^4 q} \:.
\end{equation}
The spin distribution~\eqref{S_x} is an even function of the in-plane wave vector, therefore, the spin injection via (110)-grown tunnel barriers occur even for isotropic in the interface plane distribution of incident electrons.  

Finally, we estimate the efficiency of spin injection. We consider the potential barrier of the width $a=100$\AA\, and height $V_0=200$~meV based on the GaSb/Al$_{0.17}$Ga$_{0.83}$Sb heterostructure and take the bulk GaSb parameters~\cite{Jancu2005}: $m=0.041m_0$, with $m_0$ being the free electron mass, and $\gamma=186$~eV$\cdot$\AA$^3$. The interface Rashba coefficient is estimated~\cite{Gerchikov1992,Pfeffer1999} as $\alpha \approx (P^2/3)\{\Delta_1/[(E-E_{\Gamma_8}^{(1)})(E-E_{\Gamma_7}^{(1)})] - \Delta_2/[(E-E_{\Gamma_8}^{(2)})(E-E_{\Gamma_7}^{(2)})]\}$, where $P$ is the Kane matrix element,
$E$ is the electron energy, $\Delta_j=E_{\Gamma_8}^{(j)} - E_{\Gamma_7}^{(j)}$, $E_{\Gamma_8}^{(j)}$ and $E_{\Gamma_7}^{(j)}$ are the energies of the $\Gamma_8$ and $\Gamma_7$ valence bands at $\bm k=0$ outside ($j=1$) and inside ($j=2$) the barrier. This yields $\alpha \approx 2.5$~eV$\cdot$\AA$^2$ for the Kane matrix element in GaSb $P=9.7$~eV, the band gap $0.81$~eV, the valence band offset $\approx 70$~meV and $\Delta_1=\Delta_2 \approx 0.76$~eV, the above parameters are from Refs.~\cite{Jancu2005,Kroemer2004}. Then, for the wave vectors $k_x=k_z=2\times 10^6$~cm$^{-1}$, the spin polarization $P_s = 2 |S_x|$ of electrons transmitted through the barrier can be estimated as $0.1$\%. This value is not high, however, well above the spin polarization detectable in experiments. The efficiency of spin injection can be enhanced in multiple-barrier structures provided spin relaxation processes are slow enough. 

\subsection{Spin-galvanic effect}

Besides the spin polarization induced by tunneling current through the barrier, one can consider the inverse effect, namely, the emergence of a direct electric current through the barrier in the presence of spin polarization, which is also possible in (110)-grown structures. Now we assume that the tunnel structure is initially unbiased but electrons at the both sides of the barrier are spin polarized along the $x$ axis. Due to spin-orbit coupling, the barrier transparency for the electrons incident on the barrier from the left with the initial spin polarization $S_x$ is different from that for the electrons incident of the barrier from the right. As a result, the oppositely directed electron fluxes do not compensate each other which leads to a direct electric current thought the barrier. 

Taking into account that the transmission coefficients for the electrons incident upon the barrier from the left and the right are connected to each other by the formal replacement $a \rightarrow -a$ [see discussion after Eqs.~\eqref{refl} and~\eqref{transm}], one can write for the tunneling current density
\begin{eqnarray}\label{j_tun}
j_z &=& e \sum_{\bm k} {\rm Tr} [\mathcal T_{\bm k}(a) \rho_l \mathcal T_{\bm k}^\dag(a) ] v_z \Theta(v_z) \\
&+& e \sum_{\bm k} {\rm Tr} [\mathcal T_{\bm k}(-a) \rho_r \mathcal T_{\bm k}^\dag(-a) ] v_z \Theta(-v_z) \nonumber \:,
\end{eqnarray}
where $\rho_l$ and $\rho_r$ are the electron spin-density matrices on the left-hand and right-hand sides of the structure~\cite{Tarasenko04}. For the same electron distribution at both side of the structure described by the distribution function $f(\varepsilon)$, where $\varepsilon$ is the electron energy, and small degree of spin polarization $p_s$ along the $x$ axis, the density matrices are given by
\begin{equation}\label{rho}
\rho_l = \rho_r  = f(\varepsilon) I - \frac{d f(\varepsilon)}{d \varepsilon} \frac{2 p_s \sigma_x}{\langle 1/\varepsilon \rangle} \:,
\end{equation}
where $\langle 1/\varepsilon \rangle$ is the mean value of the reciprocal kinetic energy, which is equal to $3/E_F$ for three-dimensional degenerate electron gas with the Fermi energy $E_F$ and $2/(k_B T)$ for nondegenerate gas at the temperature $T$. The calculation of the tunneling current Eq.~\eqref{j_tun} with the transmission coefficient~\eqref{Tk:res} and the density matrices~\eqref{rho} yields 
\begin{equation}\label{j:sge}
j_z = - 8 \frac{e \alpha \gamma m^2 p_s}{\langle 1/ \varepsilon \rangle \hbar^4} \sum_{\bm k} \frac{d f(\varepsilon)}{d \varepsilon} \frac{k_x^2 \, v_z}{q}  |t_{\bm k}|^2 \, {\rm Re}(c_2) \:.
\end{equation}
Finally, for the degenerate electron gas at the both sides of the barrier, the high and thick barrier, $q \gg k_F$, $q a \gg 1$, and $k_F^2 a /q \ll 1$, we derive
\begin{equation}\label{j:sge2}
j_z = \frac{64 e p_s}{105 \pi^2} \frac{\alpha \gamma m \, a k_F^9}{\hbar^3 \kappa^3}  \exp{(-2 \kappa a)}  \:,
\end{equation}
where $k_F$ is the Fermi wave vector and $\kappa = \sqrt{2 m V_0 /\hbar^2}$. The estimation following Eq.~\eqref{j:sge2} gives $j_z \sim 2$~mA/cm$^2$ for the single barrier of the width $a=100$\AA\, and height $V_0=200$~meV based on GaSb/Al$_{0.17}$Ga$_{0.83}$Sb, the Fermi wave vector $k_F = 2\times 10^6$~cm$^{-1}$, and the spin polarization $p_s=10$\%. Electric currents of a such density are reliably detectable in experiments. 

\section{Summary}

We have developed the analytical theory of electron spin injection through non-magnetic semiconductor barriers with the (110) crystallographic orientation. The calculations are carried out by using the spin-dependent transfer matrix technique.  
It is shown that the tunneling of electrons through the barrier, which are described by an isotropic momentum distribution in the interface plane, results in their partial spin polarization along the $[1\bar{1}0]$ axis. The microscopic mechanism of the spin polarization is a two-step process comprising spin filtering due to Dresselhaus spin-orbit coupling in the barrier interior and the spin rotation in the Rasha effective magnetic field at the interfaces. The inverse effect, emergence of a direct tunneling current through the barrier due to spin polarization of carriers, is also described. The spin-dependent tunneling phenomenon could by employed for designing pure electric spin injectors and detectors.   

\acknowledgements

Financial support by the Russian Foundation for Basic Research, RF President grants NSh-1085.2014.2 and MD-3098.2014.2, EU projects POLAPHEN and SPANGL4Q, and the ``Dynasty'' Foundation is gratefully acknowledged.

\appendix
\section{Effect of $\mathcal H_{D1}$ on tunneling}

The $\bm k$-cubic spin-orbit splitting of the conduction band is valid for small wave vectors and should be treated as a perturbation only. In particular, the term $\mathcal H_{D1}$ contains the operator $\partial^3 /\partial z^3$, therefore, the 
Schr\"{o}dinger equation with the Hamiltonian $\mathcal H_0 + \mathcal H_{D1}$ in the barrier would have the basic solutions $\propto \exp(q_z z) $ with three different $q_z$ for a given electron energy. One of the solutions is a rapidly oscillating function with $|q_z| \propto \gamma^{-1}$, it is unphysical and beyond of the effective Hamiltonian approximation. Besides, the operator $\mathcal H_{D1}$ is not Hermitian for the class of wave functions of heterostructures with abrupt interfaces. Therefore, the conventional procedure of calculating the transmission and reflection coefficients by constructing the basis solutions of the Schr\"{o}dinger equation in each layer of the structure and using boundary conditions at interfaces become ambiguous. Instead, we analyze the effect of spin-orbit coupling $\mathcal H_{D1}$ on tunneling in the framework of perturbation theory. Since the term $\mathcal H_{D1}$ does not contain the in-plane wave vector, we assume for simplicity that $k_x,k_y=0$ and consider one-dimensional case.  

In the absence of spin-orbit coupling, the wave functions used in the theory of electron tunneling through a rectangular barrier are given by 
\begin{equation}\label{wf}
\psi_{k}^{(+)}(z) = \left\{
\begin{array}{l}
e^{ik z} + r_{k} e^{- i k z} \:,\;\; z<0 \:,
\\
A_k e^{q z} + B_k e^{-q z}\:,\;\; 0<z<a \:,
\\
t_{k}  e^{ik (z-a)}\:,\;\; z>a \:,
\end{array} \right. 
\end{equation}
where $t_k$ and $r_k$ are the amplitude transmission and reflection coefficients [cf. Eq.~\eqref{tk:res}],
\begin{equation}\label{tr}
t_{k} = \left[ \cosh{q a} + \frac{i}{2}\left( \frac{q}{k}-
\frac{k}{q} \right) \sinh{q a} \right]^{-1} \:,
\end{equation}
\[
r_{k}= -\frac{i}{2}\left[\frac{q}{k} + \frac{k}{q}\right] \sinh{(q a)} \,
t_{k} \:,
\]
$A$ and $B$ are the function amplitudes in the barrier,
\begin{equation}
A_{k}=\frac{t_{k}}{2}\left(1+i\frac{k}{q}\right)e^{-q a} \:,\;
B_{k}=\frac{t_{k}}{2}\left(1-i\frac{k}{q}\right)e^{q a} \:,
\end{equation}
$k=\sqrt{2 m E /\hbar^2}>0$ is the initial wave vector, $q=\sqrt{2 m (V_0 -E) /\hbar^2}$, and we assume that the effective masses inside and outside the barrier coincide.
In terms of the scattering problem, the wave function $\psi_{k}^{(+)}$ at $|z| > a$ represents the sum of the incident wave and the wave diverging from the barrier. Besides the functions $\psi_{k}^{(+)}$, one can formally consider another type of the Schr\"{o}dinger equation solutions which at $|z| > a$ describe the sum of the incident and converging waves. The corresponding wave functions can be presented in the form
\begin{equation}\label{wf2}
\psi_{k}^{(-)}(z) = \left\{
\begin{array}{l}
t_k^* e^{ik z} \:,\;\; z<0 \:,
\\
B_k^* e^{q (z-a)} + A_k^* e^{-q (z-a)}\:,\;\; 0<z<a \:,
\\
 e^{i k (z-a)} + r_k^* e^{- i k (z-a)} \:,\;\; z>a \:.
\end{array} \right. 
\end{equation}

Exploiting the analogy between the theory of tunneling and the theory of quantum scattering~\cite{Migdal,LL3}, we obtain the  
first-order correction to the amplitude transmission coefficient caused by the Hermitian perturbation $\mathcal H_{D1}$, 
\begin{equation}\label{t_correction} 
\delta t_k = - \frac{i \, m}{\hbar^2 k} \, \langle \,
\psi_k^{(-)} \, | \,  \mathcal H_{D1}  \, | \, \psi_k^{(+)} \, \rangle\:.
\end{equation}
To calculate $\delta t_k$ we take into account that the functions $\psi_k^{(\pm)}$ satisfy the Schr\"{o}dinger equation with the Hamiltonian $\mathcal H_0$ and the symmetrization in Eq.~\eqref{H_D1} can be done in one of two possible ways:
\begin{subequations}
\begin{equation}
\label{way1}
\left\{ \gamma(z),\frac{\partial^3}{\partial
z^3}\right\}_{\mathrm{sym}} = \frac{1}{2} \frac{\partial}{\partial z}
\left[ \gamma(z)\frac{\partial}{\partial z}
+\frac{\partial}{\partial z} \gamma(z)\right]
\frac{\partial}{\partial z}
\end{equation}
or
\begin{equation}
\label{way2}
\left\{ \gamma(z), \frac{\partial^3}{\partial
z^3}\right\}_{\mathrm{sym}} =
\frac{1}{2} \left[ \gamma(z)\frac{\partial^3}{\partial z^3}
+\frac{\partial^3}{\partial z^3} \gamma(z)\right]
 \:.
\end{equation}
\end{subequations}
For the both types of symmetrization, the calculation yields the same result
\begin{equation}\label{t_correction2}
\delta t_k = \frac{\sigma_x m^2}{2 \hbar^4 k} \int \gamma(z) [V(z) - E] \left[ \psi_k^{(-)*} \psi_k^{{(+)}'} - \psi_k^{{(-)*}'} \psi_k^{(+)} \right] dz \:,
\end{equation}
where $\psi_k^{{(\pm)}'} = d \psi_k^{(\pm)} /dz$. In the limiting case of rectangular barrier with the spin-orbit splitting of electron states in the barrier given by $\mathcal H_{D1}$, the correction to amplitude transmission coefficient assumes the form
\begin{equation}\label{t_correction3}
\delta t_k = {i \sigma_x \over 2} {\gamma m q^2 a \over \hbar^2} t_k \:.
\end{equation}
The probability for an electron with the spin projection $s_x =\pm 1/2$ to tunnel through the barrier is given by $| \langle s_x |t_k +\delta t_k| s_x\rangle|^2$. To first order in $\gamma$, where the above consideration is valid, the probability is independent of the spin orientation since $\mathrm{Re}(t_k \,\delta t_k^*) \equiv 0$. Therefore, the $\mathcal H_{D1}$ spin-orbit coupling does not lead to a spin polarization of electrons transmitted through the barrier. However, the correction $\delta t_k$ results in the spin-dependent phase shift of the wave function which can be revealed in interference experiments. Similar conclusions on the effect of the term $\mathcal H_{D1}$ on tunneling were made in Refs.~\cite{Nguyen2009,Nguyen2009b} by considering an imaginary correction to $q$.

\end{document}